\documentclass{article}

\PassOptionsToPackage{numbers}{natbib}
 \usepackage[final]{neurips_2022}

\usepackage[utf8]{inputenc} 
\usepackage[T1]{fontenc}    
\usepackage{xr-hyper}
\usepackage{hyperref} 
\usepackage{url}            
\usepackage{booktabs}       
\usepackage{amsfonts}       
\usepackage{nicefrac}       
\usepackage{microtype}      
\usepackage{longtable}
\usepackage{adjustbox}
\usepackage{multirow}
\usepackage{graphicx}
\usepackage{todonotes}
\usepackage{colortbl}
\title{Multi-modal deep learning system for depression and anxiety detection}

\author{
  Brian Diep \\
  Department of Computer Science \\
  University of Toronto\\
  Toronto, ON \\
  \texttt{bdiep@cs.utoronto.edu}
  \thanks{This work was done while at Winterlight Labs}
  \And
  Marija Stanojevic \\
  Winterlight Labs \\
  Toronto, ON \\
  \texttt{marija@winterlightlabs.com}
  \And
  Jekaterina Novikova \\
  Winterlight Labs \\
  Toronto, ON \\
  \texttt{jekaterina@winterlightlabs.com}
}

\begin{document}

\maketitle

\begin{abstract}
Traditional screening practices for anxiety and depression pose an impediment to monitoring and treating these conditions effectively. However, recent advances in NLP and speech modelling allow textual, acoustic, and hand-crafted language-based features to jointly form the basis of future mental health screening and condition detection. Speech is a rich and readily available source of insight into an individual's cognitive state and by leveraging different aspects of speech, we can develop new digital biomarkers for depression and anxiety. To this end, we propose a multi-modal system for the screening of depression and anxiety from self-administered speech tasks. The proposed model integrates deep-learned features from audio and text, as well as hand-crafted features that are informed by clinically-validated domain knowledge. We find that augmenting hand-crafted features with deep-learned features improves our overall classification F1 score comparing to a baseline of hand-crafted features alone from 0.58 to 0.63 for depression and from 0.54 to 0.57 for anxiety. The findings of our work suggest that speech-based biomarkers for depression and anxiety hold significant promise in the future of digital health.
\end{abstract}

\section{Introduction}

Depression and anxiety are two of the most common psychiatric disorders that, depending on their severity, can have a profound impact on an individual's well-being and the quality of life \citep{henning2007impairment, gurland1992impact, roshanaei2009longitudinal,lepine2002epidemiology, richards2011prevalence}. Thus, it is imperative that treatments for depression and anxiety are prioritized as intervention can greatly improve patient outcomes \citep{dadds1997prevention, reynolds2012early}. Global improvement of anxiety and depression  treatment options is estimated to have a direct economic benefit over the period from 2016 to 2030 of \$239 billion and \$169 billion, respectively \citep{CHISHOLM2016415}.

Despite the importance of bettering the treatment pipeline, many barriers remain. One of the primary barriers to effective depression and anxiety treatment is the screening process. Traditional methods for screening have a high burden on clinicians and patients in terms of their ease of administration and scoring, no clear reference standard, and the degree of patient activation and monitoring required \citep{nease2003depression}. 
Assessment scales such as the Patient Health Questionnaire (PHQ-8) \citep{lowe2004monitoring} or Generalized Anxiety Disorder (GAD-7) \citep{spitzer2006brief} offer a more quantitative basis for screening. 

From another perspective, speech and language are two modalities that form a promising and objective basis for mental health screening. It is well-established that depression and anxiety can alter an individual's general cognition, with specific biases in their attention and memory \citep{cohen1982effort, mathews2005cognitive}. These deficits can manifest in altered acoustic and linguistic dimensions of speech. Some of these include altered rate of speech or increased usage of first-person pronouns \citep{pope1970anxiety, junghaenel2008linguistic}.

With recent advances in natural language processing and computational power, we now have the ability to collect, measure, and analyze speech data on a larger scale. There is also the rise in popularity of digital platforms such as Amazon Mechanical Turk (mTurk) that has eased the burden of data collection from clinically significant populations \citep{engle2020amazon, tasnim2022depac}. All of this has accelerated development of ML models using speech-based biomarkers for depression and anxiety. These include models that classify anxiety and depression as well as those that predict the severity of these diseases \citep{banerjee2021predicting, toto2021audibert, yang2017multimodal}. We build upon the existing literature and extend AudiBERT \citep{toto2021audibert} for the classification of depression and anxiety from speech. Our model incorporates more recent sub-module advances in the architecture and experimental settings. Importantly, we also combine both deep-learned and hand-crafted features to best capture the signal of depression and anxiety that is carried through the acoustic and linguistic properties of speech. We demonstrate that our model achieves better performance on the validation dataset.

\begin{figure}[t]
    \centering
    \includegraphics[width=1\textwidth]{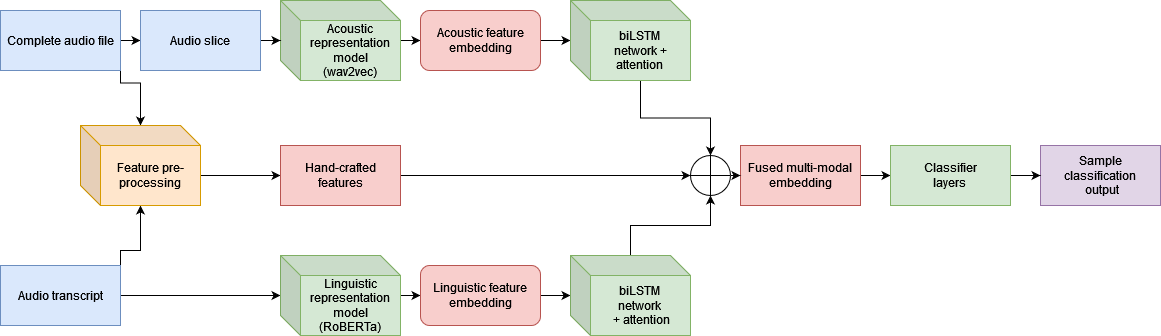}
    \caption{Classification module architecture diagram} 
    \label{fig:clf_diagram}
\end{figure}

\section{Modeling}
Depression and anxiety can present themselves through acoustic and linguistic features of speech \citep{pope1970anxiety, junghaenel2008linguistic}. Therefore, our architecture (Figure \ref{fig:clf_diagram}) leverages both of these modalities by parallel representation learning from audio and textual data in addition to representation learning from features hand-crafted by domain experts. Our architecture is inspired by AudiBERT \citep{toto2021audibert}.

Working with deep-learned representations of speech can allow for our models to capture more abstract signals in speech that can be used for better depression/anxiety detection. In our work, we use pre-trained speech and language representation models which have been shown to be effective and robust for generating representations of acoustics and text \citep{https://doi.org/10.48550/arxiv.2006.11477, liu2019roberta}.

We utilize Wav2Vec 2.0, one of the best acoustic signal representation models, to learn the features from the speech signal. The output of the Wav2Vec 2.0 base-model, pre-trained on 100k hours of the Vox-Populi dataset \citep{voxpopuli}, is forwarded to a two-layer biLSTM \citep{hochreiter1997long, graves2005framewise} and then to a multi-head attention layer with two heads. Vectors \textbf{$R1$}, outputs of multi-head attention representing acoustic signal, are used jointly with linguistic and hand-crafted features for classification.

Transformers-based architectures \citep{bert} have significantly improved language representation, and performance on variety of domain-specific tasks including emotion classification \citep{siriwardhana2020jointly}. To represent transcripts of human speech, we select the base model of RoBERTa \citep{liu2019roberta}, as one of the best performing language models which can be trained with a single GPU. The output of RoBERTa is forwarded to a two-layer biLSTM, whose output is redirected to a multi-head attention layer with two heads. Vectors \textbf{$R2$} are outputs of multi-head attention representing the linguistic signal.

There is a rich body of work studying the pathology of depression and anxiety that suggests specific changes in the acoustic, the semantic, and lexico-syntactic content of the speech of those who are suffering from these diseases \citep{pope1970anxiety, junghaenel2008linguistic}. We use domain-experts hand-crafted features \textbf{$R3$} as additional signal. List of those features can be found in the Appendix in Table \ref{tab:lin_feats} and \ref{tab:acoust-feats}. 

Vectors $R1$, $R2$, and $R3$ are concatenated together to create a combined representation embedding of the subject's speech. This representation is passed through two feedforward layers followed by a binary cross-entropy loss. The architecture classifies between disease and no disease. We train two different models, one for depression and another for anxiety task.

\section{Experimental setup}
We train and evaluate our models using 5-fold cross-validation with the folds constructed such that there is no overlap between the subjects in the training and test fold. We report the mean of precision, recall and F1-score for each model over the 5 folds. Results are achieved using AdamW optimization with learning rate $lr = 3e-5$. We use binary cross-entropy with logits loss from the PyTorch library. The model is trained on T4 Tensor Core GPU with 16 GB RAM.

We use Wav2Vec 2.0 and RoBERTa implementations from the \emph{HuggingFace} library. Due to memory and architecture constraints with inputting large audio files into Wav2Vec2, we also split the audio samples into consecutive 10 second intervals. The audio was sampled at a rate of 16 000 Hz. Then Wav2Vec2 feature extractor is used to create the input. RoBERTa's input is a speech transcript generated from the audio via ASR. The text is further transformed by the RoBERTa tokenizer and padded to length of 512. Note, we also add several tokens to the tokenizer corresponding to a set of unfilled and filled pauses in the speech. Pre-trained model weights are not frozen and are fine-tuned for 10 epochs with a batch size of 4 due to GPU memory constraints. 

As a baseline, we train a feedforward network using hand-crafted features provided by domain-experts only. The network consists of five linear layers followed by Leaky ReLU activation function \citep{maas2013rectifier}. Every layer is twice smaller than the previous one and we use a dropout of 0.2 throughout the network. Network is trained using AdamW optimization with learning rate $lr=3e-4$, batch size of 8, and binary cross-entropy with logits loss. We use the same 5-fold cross-validation process as for the proposed model and we report the mean of precision, recall and F1-score.

\subsection{Dataset}

The dataset used to train and test the model comes from an extended version of the DEPAC corpus \cite{tasnim2022depac}. The DEPAC corpus contains crowd-sourced (mTurk) audio samples from 3543 unique individuals performing a range of self-administered speech tasks. For the purposes of this analysis, we subset the data to only include speech from the tasks that contain elements of narrative speech. In total, the dataset contains 4209 unique audio samples and corresponding audio transcripts from the below-mentioned speech tasks.

\textbf{Journaling} and \textbf{prompted narrative tasks:} the participant is asked to describe an experience or event based on a given prompt. For journaling task, they are asked about their day whereas in prompted narrative, they are also asked about hobbies or travel experiences depending on the specific prompt. These narrative speech tasks can contain signals relevant for depression or anxiety prediction \citep{trifu2017linguistic}.

\textbf{Semantic fluency task:} the participant is prompted to describe within one minute positive experiences that will occur in the future. Similar verbal fluency tasks have been shown to correlate with issues with executive function associated with depression \citep{fossati2003qualitative}.

The dataset contains the self-rated PHQ-8 and GAD-7 scores for each individual. GAD-7 is rated on a scale of 0-21 and PHQ-8 on a scale of 0-24. Following AudiBERT, literature \citep{lowe2004monitoring, spitzer2006brief}, and consultations with experts, we adopt binary classification tasks. We convert these scores into a "soft" binary diagnosis label using a score of 10 as a cutoff on both scales. Approximately 25.3\% of subjects had a PHQ-8 score above 9, and 12.8\% had a GAD-7 score above 9 (diagnosis).

For each complete audio sample and transcript, we extract the hand-crafted features whose list is given in the Appendix \ref{tab:lin_feats} and \ref{tab:acoust-feats}.

\section{Results and Discussion}

The results of our experiments are displayed in Table \ref{tab:clf_results}.
Examining them in aggregate reveals that our models perform better in predicting no diagnosis, and they are struggling to predict diagnosis. We hypothesize that this is partially a function of the data imbalance that exists within our dataset, as most collected depression and anxiety data comes from individuals with lower scores.

 The results show that the inclusion of deep-learned features enriches the representation by adding properties that are not fully captured the hand-crafted features, improving the detection of depression and anxiety. This reflects previous results \citep{toto2021audibert}, where the addition of deep-learned features, especially text representation models, improved classification performance for depression.

One of the challenges with developing models for classification of depression and anxiety comes from the distribution of data. In our data and much corpora, a majority of the subjects ware classified with having PHQ-8/GAD-7 scores under 10 leading to class imbalance \cite{valstar2014avec, gratch2014distress}. Imbalance in classes in training data poses a hurdle in development of robust models \citep{krawczyk2016learning}. Furthermore, within the classes, the distribution of scores is still uneven. A distribution of PHQ-8/GAD-7 scores is long-tailed and skewed towards lower severity cases. This can lead to issues of within-class imbalance that are difficult to resolve \citep{japkowicz2001concept}.

\begin{table}
\caption{Anxiety and depression classification results. Bold indicates highest F1 score per disease.}
\begin{adjustbox}{max width=1\linewidth, center}
\begin{tabular}{|l|lll|lll|lll|lll|}
\hline
 & \multicolumn{6}{c|}{Anxiety} & \multicolumn{6}{c|}{Depression} \\ \cline{2-13}
 & \multicolumn{3}{c|}{\begin{tabular}[c]{@{}l@{}}Hand-crafted\\ features only\end{tabular}} & \multicolumn{3}{l|}{\begin{tabular}[c]{@{}c@{}}Deep-learned + hand-crafted\\ features\end{tabular}} & \multicolumn{3}{c|}{\begin{tabular}[c]{@{}c@{}}Hand-crafted\\ features only\end{tabular}} & \multicolumn{3}{l|}{\begin{tabular}[c]{@{}c@{}}Deep-learned + hand-crafted\\ features\end{tabular}} \\ \cline{2-13}
& \multicolumn{1}{l|}{Precision} & \multicolumn{1}{l|}{Recall} & F1 & \multicolumn{1}{l|}{Precision} & \multicolumn{1}{l|}{Recall} & F1 & \multicolumn{1}{l|}{Precision} & \multicolumn{1}{l|}{Recall} & F1 & \multicolumn{1}{l|}{Precision} & \multicolumn{1}{l|}{Recall} & F1 \\ \hline
\begin{tabular}[c]{@{}l@{}}No diagnosis\\ ( score\textless 10)\end{tabular} & \multicolumn{1}{l|}{0.81} & \multicolumn{1}{l|}{0.65} & 0.72 & \multicolumn{1}{l|}{0.76} & \multicolumn{1}{l|}{0.72} & \textbf{0.73} & \multicolumn{1}{l|}{0.73} & \multicolumn{1}{l|}{0.78} & 0.75 & \multicolumn{1}{l|}{0.77} & \multicolumn{1}{l|}{0.83} & \textbf{0.80}\\ \hline
\begin{tabular}[c]{@{}l@{}}Diagnosis\\ (score$\geq$ 10)\end{tabular} & \multicolumn{1}{l|}{0.28} & \multicolumn{1}{l|}{0.41} & 0.33 & \multicolumn{1}{l|}{0.37} & \multicolumn{1}{l|}{0.42} & \textbf{0.40} & \multicolumn{1}{l|}{0.31} & \multicolumn{1}{l|}{0.42} & 0.35 & \multicolumn{1}{l|}{0.48} & \multicolumn{1}{l|}{0.39} & \textbf{0.43}\\ \hline
Overall & \multicolumn{2}{l|}{\cellcolor[HTML]{656565}} & 0.54 & \multicolumn{2}{l|}{\cellcolor[HTML]{656565}} & \textbf{0.57} & \multicolumn{2}{l|}{\cellcolor[HTML]{656565}} & 0.58 & \multicolumn{2}{l|}{\cellcolor[HTML]{656565}{\color[HTML]{656565} }} & \textbf{0.63} \\ \hline
\end{tabular}
\end{adjustbox}
\label{tab:clf_results}
\end{table}

Interestingly, we also find that depression classification results in higher overall F1-score than anxiety classification. One reason for this was likely due to the data imbalance issue in anxiety samples, which was particularly pronounced as compared to depression (12.8\% vs. 25.3\% with scores above 9). Another potential reason for this worse performance is that acoustic features in anxiety have been shown to not vary as much with severity as compared with depression \citep{albuquerque2021association}. This suggests that anxiety prediction through speech-assessment is a harder task than its corollary in depression. 

These findings add to the existing body of work that speech is an appropriate modality for depression and anxiety biomarker development. In particular, using both hand-crafted and deep-learned features maximizes the signal that can be extracted from the speech stream. It also shows how prediction performance for these models is often variable with respect to anxiety/depression severity.

\section{Conclusion}
In this work, we present a model for the prediction of anxiety and depression from self-administered speech tasks. Our models extend upon previous work that focuses on classification of depression and anxiety and combines it with a set of hand-crafted features that is able to capture many of the nuanced changes in acoustic and linguistic content of depressed and anxious speech.
We find that the proposed model, that combines hand-crafted features with deep-learning speech and language representation, improves classification F1-score of both classes compared to the baseline. 
The results presented in this paper form a promising basis towards the development of better screening tools for anxiety and depression via speech data.

\newpage

\bibliographystyle{plainnat}
\bibliography{custom}

\newpage

\appendix
\counterwithin{table}{section}
\section{Appendix}\label{appendix:sup}
\renewcommand{\thetable}{\Alph{section}.\arabic{table}}

\begin{center}
\begin{table*}[htbp]
\caption{Summary of the linguistic features.}
\label{tab:lin_feats}
\renewcommand{\arraystretch}{1.3}
    \centering
    \setlength\tabcolsep{2pt}
    \begin{tabular}{|p{0.25\linewidth}|p{0.7\linewidth}|}
    \multicolumn{2}{c}{}\\
    \hline
    \textbf{Feature Group} &  \textbf{Motivations}\\
    \hline
    Discourse mapping & Techniques to formally quantify utterance similarity and disordered speech via distance metrics or graph-based representations.\\
    \hline
    Local coherence & Coherence and cohesion in speech is associated with the ability to sustain attention and executive functions.\\
    \hline
    Lexical complexity and richness & Language pattern changes in particular related to the irregular usage patterns of words of certain grammatical categories.\\
    \hline

    Syntactic complexity & Measures of syntactic complexity of utterances.\\

    \hline
    Utterance cohesion &  Measures of tense and concordance within utterances. \\
    \hline

    Sentiment & Features such as valence, arousal, and dominance. \\
    \hline
    Word finding difficulty & Metrics related to disfluency and filled pauses in speech. \\
    \hline
    \end{tabular}
\end{table*}
\end{center}

\begin{center}
\begin{table*}[htbp]
\caption{Summary of the acoustic features.}
\label{tab:acoust-feats}
\renewcommand{\arraystretch}{1.3}
    \centering
    \setlength\tabcolsep{2pt}
    \begin{tabular}{|p{0.25\linewidth}|p{0.7\linewidth}|}
    \multicolumn{2}{c}{}\\
    \hline
    \textbf{Feature Group} &  \textbf{Motivations}\\
    \hline
    Intensity (auditory model based) & Perceived loudness in $dB$ relative to normative human auditory threshold. \\
    \hline
    MFCC 0-12 &  MFCC 0-12 and energy, their first and second order derivatives are calculated on every 16 ms window and step size of 8 ms, and then, averaged over the entire sample. \\
    \hline
    Zero-crossing rate (ZCR) & Zero-crossing rate across all the voiced frames showing how intensely the voice was uttered. \\
    \hline
    $F_0$ & Fundamental frequency in Hz. \\
    \hline
    Harmonics-to-noise ratio (HNR) & Degree of acoustic periodicity. \\
    \hline
    Jitter and shimmer & Jitter is the period perturbation quotient and shimmer is the amplitude perturbation quotient representing the variations in the fundamental frequency. \\
    \hline
    Durational features & Total audio and speech duration in the sample. \\
    \hline
    Pauses and fillers & Number and duration of short ($< 1s$), medium ($1-2 s$)  and long ($>2 s$) pauses, mean pause duration, and pause-to-speech ratio. \\
    \hline
    Phonation rate & Number of voiced time windows over the total number of time windows in a sample.\\
    \hline
    \end{tabular}
\end{table*}
\end{center}

\end{document}